\documentclass[12pt,twoside,a4paper]{article}
\usepackage[T1]{fontenc}
\usepackage[utf8]{inputenc}
\usepackage{amsfonts}
\usepackage{verbatim}
\usepackage{graphicx}
\usepackage{slashed}

\usepackage{cite}
\usepackage{cancel}
\usepackage{caption}
\usepackage{subcaption}
\usepackage{epstopdf}
\usepackage{amsthm}
\usepackage{amsmath}
\usepackage{color}
\usepackage{hyperref}

\usepackage{booktabs}  
\usepackage{siunitx}

\DeclareGraphicsExtensions{.eps}

\usepackage{float}
\usepackage{numprint}

\newcommand{\newc}{\newcommand*}

\long\def\begincomment#1\endcomment{%
        \begingroup\sf\baselineskip12pt#1\endgroup}

\newc{\etal}{\textrm{et al.}} 
\newc{\eg}{\textrm{e.g.}} 
\newc{\ie}{\textrm{i.e.}}
\newc{\etc}{\textrm{etc}}
\newc\vs{\textrm{vs.}}
\newc{\cl}{\rm {C.L.}}

\newc{\ev}{\ensuremath{\,\mathrm{eV}}}
\newc{\kev}{\ensuremath{\,\mathrm{keV}}}
\newc{\mev}{\ensuremath{\,\mathrm{MeV}}}
\newc{\gev}{\ensuremath{\,\mathrm{GeV}}}
\newc{\tev}{\ensuremath{\,\mathrm{TeV}}}
\newc{\MeV}{\mev} 
\newc{\TeV}{\tev}
\newc{\invpb}{\ensuremath{/\text{pb}}}
\newc{\invfb}{\ensuremath{/\text{fb}}}
\newc\nb{\ensuremath{\,\mathrm{nb}}} \newc\pb{\ensuremath{\,\mathrm{pb}}} \newc\fb{\ensuremath{\,\mathrm{fb}}}
\newc\pc{\ensuremath{\,\mathrm{pc}}}
\newc\kpc{\ensuremath{\,\mathrm{kpc}}}
\newc\mpc{\ensuremath{\,\mathrm{Mpc}}}
\newc\ps{\ensuremath{\,\mathrm{ps}}} 
\newc\cmeter{\ensuremath{\,\mathrm{cm}}} 
\newc\meter{\ensuremath{\,\mathrm{m}}} 
\newc\kmeter{\ensuremath{\,\mathrm{km}}}
\newc\second{\ensuremath{\,\mathrm{s}}}
\newc\msecond{\ensuremath{\,\mathrm{ms}}}
\newc\nsecond{\ensuremath{\,\mathrm{ns}}}
\newc\psecond{\ensuremath{\,\mathrm{ps}}}

\newc{\chisqmin}{\ensuremath{\chi^2_{\mathrm{min}}}}
\newc{\Delchisq}{\ensuremath{\Delta\chi^2}}
\newc{\chisq}{\ensuremath{\chi^2}}
\newc{\like}{\ensuremath{\mathcal{L}}}

\newc\lsim{\ensuremath{\mathrel{\rlap{\lower4pt\hbox{\hskip1pt$\sim$}}\raise1pt\hbox{$<$}}}}
\newc\gsim{\ensuremath{\mathrel{\rlap{\lower4pt\hbox{\hskip1pt$\sim$}}\raise1pt\hbox{$>$}}}}
\newc{\VEV}[1]{\ensuremath{\langle #1 \rangle}}
\newc{\dl}{\ensuremath{\stackrel{\leftarrow}{D}}}
\newc{\dr}{\ensuremath{\stackrel{\rightarrow}{D}}}

\newc{\bcenter}{\begin{center}}    \newc{\ecenter}{\end{center}}
\newc{\bfl}{\begin{flushleft}}    \newc{\efl}{\end{flushleft}}
\newc{\bfr}{\begin{flushright}}    \newc{\efr}{\end{flushright}}

\newc{\bi}{\begin{itemize}}
\newc{\ei}{\end{itemize}}
\newc{\bed}{\begin{description}}
\newc{\eed}{\end{description}}
\newc{\ben}{\begin{enumerate}}
\newc{\een}{\end{enumerate}}

\newc{\be}{\begin{equation}}
\newc{\ee}{\end{equation}}
\newc{\bea}{\begin{eqnarray}}
\newc{\eea}{\end{eqnarray}}
\newc{\bfle}{\begin{flalign}}
\newc{\efle}{\end{flalign}}
\newc{\ra}{\rightarrow}

\newc{\alphas}{\ensuremath{\alpha_s}}
\newc{\alphatwo}{\ensuremath{\alpha_2}}
\newc{\alphaone}{\ensuremath{\alpha_1}}
\newc{\alphai}[1]{\ensuremath{\alpha_{#1}}}
\newc{\alphaem}{\ensuremath{\alpha_{\mathrm{em}}}}
\newc{\alphaeff}{\ensuremath{\alpha_{\mathrm{eff}}}}
\newc{\sineff}{\ensuremath{\sin^2 \theta_{\mathrm{eff}}}}
\newc{\sinsqeff}{\ensuremath{\sin^2 \theta_{\mathrm{eff}}}}
\newc{\dalphahad}{\ensuremath{\Delta \alpha_{\mathrm{had}}}}
\newc{\yt}{\ensuremath{h_t}} \newc{\yb}{\ensuremath{h_b}} \newc{\ytau}{\ensuremath{h_{\tau}}}
\newc\mz{\ensuremath{m_Z}} 
\newc\mw{\ensuremath{m_W}}
\newc\mZ{\mz}        \newc\mW{\mw}
\newc\mhsm{\ensuremath{ m_{H_{\mathrm{SM}}}}}
\newc{\mtop}{\ensuremath{ m_t}}               
\newc{\mbottom}{\ensuremath{ m_b}} 
\newc{\mtau}{\ensuremath{ m_{\tau}}}
\newc{\mt}{\mtpole}
\newc{\mb}{\mbottom} 
\newc{\rtwogg}{\ensuremath{R_{h_2}(\gamma\gamma)}}
\newc{\rtwozz}{\ensuremath{R_{h_2}(ZZ)}}
\newc{\ronegg}{\ensuremath{R_{h_1}(\gamma\gamma)}}
\newc{\ronezz}{\ensuremath{R_{h_1}(ZZ)}}
\newc{\rsiggg}{\ensuremath{R_{h_\textrm{sig}}(\gamma\gamma)}}
\newc{\rsigzz}{\ensuremath{R_{h_\textrm{sig}}(ZZ)}}
\newc{\llbar}{\ensuremath{\ell\bar{\ell}}}
\newc{\tauptaum}{\ensuremath{ \tau^+\tau^-}}
\newc{\qqbar}{\ensuremath{ q\bar{q}}} \newc{\ppbar}{\ensuremath{ p\bar{p}}}
\newc{\bbbar}{\ensuremath{ b\bar{b}}} \newc{\ttbar}{\ensuremath{ t\bar{t}}}
\newc{\ffbar}{\ensuremath{ f\bar{f}}} \newc{\tautaubar}{\ensuremath{ \tau\bar{\tau}}}

\newc{\mchi}{\ensuremath{m_\neutone}}
\newc{\squark}{\ensuremath{\tilde{q}}}
\newc{\slepton}{\ensuremath{\tilde{l}}}
\newc{\gluino}{\ensuremath{\tilde{g}}} 
\newc{\mgluino}{\ensuremath{{m_{\gluino}}}}
\newc{\wino}{\ensuremath{\tilde{W}}} 
\newc{\mwino}{\ensuremath{{m_{\wino}}}}
\newc{\tone}{\ensuremath{{\tilde{t}_1}}}
\newc{\Hone}{\ensuremath{{\tilde{H}_{1}}}}
\newc{\Htwo}{\ensuremath{{\tilde{H}_{2}}}}
\newc{\Hhtwo}{\ensuremath{{H_{2}}}}
\newc{\qli}{\ensuremath{{\tilde{Q}_{i}}}}
\newc{\uri}{\ensuremath{{\tilde{u}_{i}}}}
\newc{\dri}{\ensuremath{{\tilde{d}_{i}}}}
\newc{\lli}{\ensuremath{{\tilde{L}_{i}}}}
\newc{\eri}{\ensuremath{{\tilde{e}_{i}}}}

\newc{\sthw}{\ensuremath{ \sin\theta_W}}              \newc{\cthw}{\ensuremath{\cos\theta_W}}
\newc{\tanthw}{\ensuremath{ \tan\theta_W}}              \newc{\cotthw}{\ensuremath{\cot\theta_W}}
\newc{\ssqthw}{\ensuremath{\sin^2 \theta_W}}
\newc{\msbar}{\ensuremath{\overline{MS}}} \newc{\drbar}{\ensuremath{\overline{DR}}}
\newc{\mtmtsmmsbar}{\ensuremath{ m_t(m_t)^{\msbar}_{{\mathrm{SM}}}}}
\newc{\mtmtsmdrbar}{\ensuremath{ m_t(m_t)^{\drbar}_{{\mathrm{SM}}}}}
\newc{\mtmtmssmdrbar}{\ensuremath{ m_t(m_t)^{\drbar}_{{\mathrm{SUSY}}}}}
\newc{\mbmbmsbar}{\ensuremath{ m_b^{\msbar}(m_b)}}
\newc{\mcmbmsbar}{\ensuremath{ m_c^{\msbar}(m_c)}}
\newc{\msmbmsbar}{\ensuremath{ m_s^{\msbar}}}
\newc{\mdmbmsbar}{\ensuremath{ m_d^{\msbar}}}
\newc{\mumbmsbar}{\ensuremath{ m_u^{\msbar}}}
\newc{\mtaupole}{\ensuremath{m_\tau^{\rm pole}}}
\newc{\mmupole}{\ensuremath{m_\mu^{\rm pole}}}
\newc{\mepole}{\ensuremath{m_e^{\rm pole}}}
\newc{\mzpole}{\ensuremath{M_Z^{\rm pole}}}
\newc{\mbmbsmmsbar}{\ensuremath{ m_b(m_b)^{\msbar}_{{\mathrm{SM}}}}}
\newc{\mbmzsmmsbar}{\ensuremath{ m_b(\mz)^{\msbar}_{{\mathrm{SM}}}}}
\newc{\mbmzsmdrbar}{\ensuremath{ m_b(\mz)^{\drbar}_{{\mathrm{SM}}}}}
\newc{\mbmzmssmdrbar}{\ensuremath{ m_b(\mz)^{\drbar}_{{\mathrm{SUSY}}}}}
\newc{\mtaumzsmmsbar}{\ensuremath{ m_{\tau}(\mz)^{\msbar}_{{\mathrm{SM}}}}}
\newc{\mtaumzsmdrbar}{\ensuremath{ m_{\tau}(\mz)^{\drbar}_{{\mathrm{SM}}}}}
\newc{\mtaumzmssmdrbar}{\ensuremath{ m_{\tau}(\mz)^{\drbar}_{{\mathrm{SUSY}}}}}
\newc{\alphasmzms}{\ensuremath{\alpha_s^{\overline{MS}}(M_Z)}}
\newc{\alphaimzms}[1]{\ensuremath{\alpha_{#1}(M_Z)^{\overline{MS}}}}
\newc{\alphaemmz}{\ensuremath{\alpha_{\mathrm{em}}^{-1}(M_Z)}}

\newc{\mzero}{\ensuremath{{m_0}}}
\newc{\mhalf}{\ensuremath{ m_{1/2}}}
\newc{\tanb}{\ensuremath{\tan\beta}}
\newc{\azero}{\ensuremath{ A_0}}
\newc{\signmu}{\ensuremath{\rm{sgn}\,\mu}}
\newc{\atau}{\ensuremath{{A_{\tau}}}}
\newc{\mueff}{\ensuremath{\mu_{\rm{eff}}}}
\newc{\lam}{\ensuremath{{\lambda}}}
\newc{\kap}{\ensuremath{{\kappa}}}
\newc{\alam}{\ensuremath{{A_{\lambda}}}}
\newc{\akap}{\ensuremath{{A_{\kappa}}}}
\newc{\hs}{\ensuremath{ H_s}}      
\newc{\mhs}{\ensuremath{ m_{H_s}}} 
\newc{\mgut}{\ensuremath{ M_{\rm GUT}} }
\newc{\gut}{\ensuremath{{\rm GUT}}}
\newc{\mplanck}{\ensuremath{ M_{\rm P}}}      \newc{\mpl}{\ensuremath{ M_{\rm Pl}}}
\newc{\msusy}{\ensuremath{ M_{\rm SUSY}}}      \newc{\ms}{\ensuremath{ M_{\rm S}}}
\newc{\mew}{\ensuremath{ M_{\rm EW}}}  
 \newc{\hu}{\ensuremath{ H_u}}       \newc{\hd}{\ensuremath{ H_d}}
 \newc{\mhu}{\ensuremath{ m_{H_u}}}       \newc{\mhd}{\ensuremath{ m_{H_d}}}
 \newc{\mhuew}{\ensuremath{ m^{\ast}_{H_u}}}       \newc{\mhdew}{\ensuremath{ m^{\ast}_{H_d}}}
 \newc{\mhuewsq}{\ensuremath{ m^{\ast\, 2}_{H_u}}}       \newc{\mhdewsq}{\ensuremath{ m^{\ast\, 2}_{H_d}}}
 \newc{\mhl}{\ensuremath{m_\hl}} 
 \newc{\mhone}{\ensuremath{m_{h_1}}} 
 \newc{\mhtwo}{\ensuremath{m_{h_2}}} 
 \newc{\mhi}{\ensuremath{m_{\tilde{h}}}} 
 \newc{\mul}{\ensuremath{m_{\tilde{u}_L}}} 
 \newc{\mtone}{\ensuremath{m_{\tilde{t}_1}}} 
 \newc{\ma}{\ensuremath{m_A}} 
 \newc{\mH}{\ensuremath{m_H}} 
 \newc{\maone}{\ensuremath{m_{a_1}}} 
 \newc{\matwo}{\ensuremath{m_{a_2}}}
 \newc{\hone}{\ensuremath{h_1}}
 \newc{\htwo}{\ensuremath{h_2}}
 \newc{\aone}{\ensuremath{a_1}}
 \newc{\atwo}{\ensuremath{a_2}}
 \newc{\mqthree}{\ensuremath{m_{\tilde{q}_3}^2}}
 \newc{\muthree}{\ensuremath{m_{\tilde{u}_3}^2}}
 \newc{\mql}{\ensuremath{m_{\tilde{q}}}}
 \newc{\mqlij}{\ensuremath{(m^{\tilde{q}}_{ij})^2}}
 \newc{\mur}{\ensuremath{m_{\tilde{u}}}}
 \newc{\mdr}{\ensuremath{m_{\tilde{d}}}}
 \newc{\murij}{\ensuremath{(m^{\tilde{u}}_{ij})^2}}
 \newc{\md}{\ensuremath{m_{\tilde{D}}}}
 \newc{\me}{\ensuremath{m_{\tilde{E}}}}
 \newc{\muu}{\ensuremath{m_{\tilde{U}}}}
 \newc{\mdrij}{\ensuremath{(m^{\tilde{d}}_{ij})^2}}
 \newc{\mll}{\ensuremath{m_{\tilde{l}}}}
 \newc{\mllij}{\ensuremath{(m^{\tilde{l}}_{ij})^2}}
 \newc{\mdlij}{\ensuremath{(m^{dl}_{ij})^2}}
 \newc{\mer}{\ensuremath{m_{\tilde{e}}}}
 \newc{\merij}{\ensuremath{(m^{\tilde{e}}_{ij})^2}}
 \newc{\ts}{\ensuremath{T_{SUSY}}}

\newc{\sigsip}{\ensuremath{\sigma^{\rm SI}_{p}}}	\newc{\sigsin}{\ensuremath{\sigma^{\rm SI}_{n}}}
\newc{\sigsdp}{\ensuremath{\sigma^{\rm SD}_{p}}}	\newc{\sigsdn}{\ensuremath{\sigma^{\rm SD}_{n}}}
\newc{\sigsi}{\ensuremath{\sigma^{\rm SI}}}	\newc{\sigsd}{\ensuremath{\sigma^{\rm SD}}}
\newc{\abund}{\ensuremath{ \Omega h^2}}
\newc{\omegadm}{\ensuremath{ \Omega_{{\rm DM}}}}     \newc{\abunddm}{\ensuremath{ \Omega_{{\rm DM}} h^2}} 
\newc{\omegam}{\ensuremath{ \Omega_{{\rm m}}}}       \newc{\abundm}{\ensuremath{ \Omega_{{\rm m}} h^2}}
\newc{\omegab}{\ensuremath{ \Omega_{{\rm b}}}}	\newc{\abundb}{\ensuremath{ \Omega_{{\rm b}} h^2}}
\newc{\omegatot}{\ensuremath{ \Omega_{{\rm TOT}}}}
\newc{\omegacdm}{\ensuremath{ \Omega_{{\rm CDM}}}}   \newc{\abundcdm}{\ensuremath{ \Omega_{{\rm CDM}} h^2}}
\newc{\omegalambda}{\ensuremath{ \Omega_{\Lambda}}} \newc{\abundlambda}{\ensuremath{ \Omega_{\Lambda} h^2}}
\newc{\omegarad}{\ensuremath{ \Omega_{{\rm rad}}}}  \newc{\abundrad}{\ensuremath{ \Omega_{{\rm rad}} h^2}}
\newc{\rhocrit}{\ensuremath{ \rho_{\rm crit}}}
\newc{\rhochi}{\ensuremath{ \rho_{\chi}}}
\newc{\abunchi}{\ensuremath{\Omega_\chi h^2}}
\newc{\abundlsp}{\ensuremath{\Omega_{\rm LSP}h^2}}

\newc{\amu}{\ensuremath{ a_{\mu}}}        \newc{\amususy}{\ensuremath{ a_{\mu}^{\mathrm{SUSY}}}}
\newc{\amuexpt}{\ensuremath{ a_{\mu}^{\mathrm{expt}}}}        \newc{\amusm}{\ensuremath{ a_{\mu}^{\mathrm{SM}}}}
\newc\deltaamu{\ensuremath{\Delta a_{\mu}}} \newc{\deltaamususy}{\ensuremath{\delta a_{\mu}^{\mathrm{SUSY}}}}
\newc\gmtwo{\ensuremath{ (g-2)_{\mu}}} 
\newc{\deltagmtwomususy}{\ensuremath{\delta\left(g-2\right)_{\mu}^{\mathrm{SUSY}}}}
\newc{\deltagmtwomu}{\ensuremath{\delta\left(g-2\right)_{\mu}}}
\newc\BR{\ensuremath{\rm BR}}
\newc\bsgamma{\ensuremath{ b\rightarrow s \gamma }}
\newc\bxsgamma{\ensuremath{\overline{B}\rightarrow X_{s}\gamma}}
\newc\brbsgamma{\ensuremath{\BR\left(\bsgamma\right)}}
\newc\brbxsgamma{\ensuremath{\BR\left(\bxsgamma\right)}}
\newc\bsmumu{\ensuremath{B_s\to\mu^+\mu^-}}
\newc\bdmumu{\ensuremath{B_d\to\mu^+\mu^-}}
\newc\brbsmumu{\ensuremath{\BR\left(B_s\to\mu^+\mu^-\right)}}
\newc\brbdmumu{\ensuremath{\BR\left(B_d\to\mu^+\mu^-\right)}}
\newc\bdmmumu{\ensuremath{\overline{B}_d\to\mu^+\mu^-}}
\newc\bbbarmix{\ensuremath{\overline{B}_s\mbox{-}B_s}}      
\newc\delmbs{\ensuremath{\Delta M_{B_s}}}
\newc\thc{\ensuremath{t\to h c}}
\newc\thu{\ensuremath{t\to h u}}
\newc{\butaunu}{\ensuremath{B_u \rightarrow \tau \nu}}
\newc{\brbutaunu}{\ensuremath{\BR\left(B_u \rightarrow \tau \nu\right)}}

\newcommand*{\neutone}{\ensuremath{\tilde{\chi}^0_1}}

\newcommand*{\multinest}{MultiNest}

\let\oldcite\cite
\renewcommand*{\cite}{~\oldcite}

\newcommand*{\hl}{\ensuremath{h}}

\newc{\Yb}{\ensuremath{Y_b}}
\newc{\Ys}{\ensuremath{Y_s}}
\newc{\Ym}{\ensuremath{Y_{\mu}}} 
\newc{\mtpole}{\ensuremath{m_t^{\rm pole}}}

\newcommand{\SidePlotsTwo}[4]
{
\begin{figure}[htbp]
\centering
\begin{subfigure}{.55\textwidth}
  \centering
  \includegraphics[height=#4]{#1}
\end{subfigure}%
\begin{subfigure}{.55\textwidth}
  \centering
  \includegraphics[height=#4]{#2}
  \end{subfigure}%
\caption{#3}
\end{figure}
}

\newcommand{\SidePlotsTwoW}[3]
{
\begin{figure}[htbp]
\centering
\begin{subfigure}{.55\textwidth}
  \centering
  \includegraphics[width=\textwidth]{#1}
\end{subfigure}%
\begin{subfigure}{.55\textwidth}
  \centering
  \includegraphics[width=\textwidth]{#2}
  \end{subfigure}%
\caption{#3}
\end{figure}
}

\newcommand{\SidePlotsTwot}[4]
{
\begin{figure}[t]
\centering
\begin{subfigure}{.55\textwidth}
  \centering
  \includegraphics[height=#4]{#1}
\end{subfigure}%
\begin{subfigure}{.55\textwidth}
  \centering
  \includegraphics[height=#4]{#2}
  \end{subfigure}%
\caption{#3}
\end{figure}
}

\begin{document}
\begin{titlepage}
\vspace*{2cm}
\begin{center} 
{\bf\Large Effects of supersymmetric threshold\\[2mm]
           corrections on the Yukawa matrix unification}\\[1cm]
{\large Mateusz Iskrzy\'nski}\\[5mm]
{\it Institute of Theoretical Physics, University of Warsaw,\\
         Pasteura 5, PL-02-093 Warsaw, Poland.}\\[3cm]
{\bf Abstract}\\[5mm] 
\end{center}
We present an updated analysis of the Yukawa matrix unification within the
renormalizable  R-parity-conserving  Minimal Supersymmetric Standard Model. It is assumed
that the soft terms are non-universal but flavour-diagonal in the super-CKM
basis at the GUT scale. Trilinear Higgs-squark-squark $A$-terms can 
generate large threshold corrections to the Yukawa matrix $\mathbf{Y}^d$
at the superpartner decoupling scale. In effect, the $SU(5)$ boundary
condition $\mathbf{Y}^d=\mathbf{Y}^{e\,T}$ at the GUT scale can be
satisfied. However, such large trilinear terms make the usual Higgs vacuum
metastable (though long-lived). We broaden previous studies by including
results from the first LHC phase, notably the measurement of the Higgs
particle mass, as well as a quantitative investigation of flavour
observables.
\end{titlepage}
\setcounter{page}{1}
\addtolength{\voffset}{-20pt}
\addtolength{\oddsidemargin}{-40pt}
\addtolength{\evensidemargin}{-40pt}
\section{Introduction}

Supersymmetric Grand Unified Theories (SUSY GUTs) have been a topic of
multiple studies since the original formulation of the $SU(5)$ 
model~\cite{Dimopoulos:1981zb}. A successful unification of gauge
couplings in the Minimal Supersymmetric Standard Model (MSSM) is 
a phenomenological triumph of this programme. GUT symmetries are
decisively helpful as providers of boundary conditions at the high  energy  scale, 
which reduces  the  dimensionality of the MSSM huge parameter space.

Despite a notable historical success of an approximate bottom-tau Yukawa
unification, the absence of such a unification for the first two
generations remains a long-standing issue. Modifications of the
GUT field content that aimed at solving this problem have
been applied already in the the non-supersymmetric
case~\cite{Georgi:1979df}. The most exhaustively studied alterations to the
boundary conditions at the GUT scale arise from assuming non-negligible
effects from higher-dimensional operators in the $SU(5)$ model
Lagrangian. Different mass ratios obtained by such manipulations are
reviewed, e.g., in Refs.~\cite{Antusch:2009gu,Antusch:2013rxa}.

An important problem of the minimal $SU(5)$ model~\cite{Dimopoulos:1981zb}
is the proton decay triggered by the higgsino triplet exchange.
Although it remains a non-trivial constraint, several analyses have shown that
the theory has not yet been excluded. The proton lifetime can be enhanced by
several orders of magnitude by contributions from higher-dimensional
operators~\cite{EmmanuelCosta:2003pu}. Moreover, it has been shown that
the tension with experimental results becomes weaker when
one uses three-loop Renormalization Group Equations (RGEs) and
two-loop decoupling conditions~\cite{Martens:2010nm}. In the present
paper, we do not restrict ourselves to the minimal model. In particular, we
allow the higgsino triplets to acquire superheavy masses from their couplings
to additional {\boldmath $5$} and {\boldmath $\bar 5$} fields that do not
couple to ordinary matter.  No proton decay problem occurs within such a
setup, while the Yukawa unification constraint remains the same as in the
minimal case.

It has been observed a long time ago that threshold corrections at the
superpartner decoupling scale $\mu_{\rm sp}$ can significantly change or even
generate the light fermion masses~\cite{Buchmuller:1982ye}. This mechanism was
applied in the context of grand unification in
Ref.~\cite{Hall:1985dx}. However, in most of the contemporary phenomenological
analyses, Yukawa unification has been exhaustively studied only in the third
generation case.

A quantitative study that achieved $Y_s(M_{\rm GUT})=Y_{\mu}(M_{\rm GUT})$
within renormalizable R-parity-conserving MSSM was performed in
Ref.~\cite{DiazCruz:2000mn}. It included only the threshold corrections coming
from gluino and higgsino loops, and concluded that a tension arises between
the Yukawa unification and flavour observables. That was likely to happen
because flavour off-diagonal soft terms were used to generate the Cabibbo
angle as well. This analysis was later broadened and simplified to the
flavour-diagonal case in Ref.~\cite{Enkhbat:2009jt}. It provided examples of
points in the MSSM parameter space where the $SU(5)$ Yukawa unification was
achieved for $\tan\beta \leq 20$. In another article~\cite{Bajc:2013dea} where
the leading MSSM threshold corrections were investigated, the problem of
proton decay was addressed by raising the Higgs soft masses above
30$\,$TeV. Examples of including such corrections in non-$SU(5)$ models can be
found, e.g., in Refs.~\cite{Aulakh:2008sn,Aulakh:2013lxa}.
 
Our work updates the one of Ref.~\cite{Enkhbat:2009jt} with a broader range of
$\tan\beta$ (reaching 40), inclusion of the contemporary experimental
data, as well as a quantitative study of flavour observables. Results from the
first phase of the LHC have constrained the superpartner masses and delivered
the lightest Higgs mass measurement, thus calling for an up-to-date analysis
of Yukawa unification.

We shall make use of a corrected account for chirally enhanced threshold
corrections to fermion masses in the MSSM. It was summarized in
Ref.~\cite{Crivellin:2011jt} and earlier published as parts of other
analyses~\cite{Crivellin:2008mq,Crivellin:2010er,Crivellin:2010gw}.  However,
the two-loop effects computed and described in Ref.~\cite{Crivellin:2012zz}
are not included in the present work.  It might be interesting to study their
effect in the future even though they are unlikely to affect our final
conclusions.

The article is organised as follows. In Sec.~\ref{ModelSec}, our
phenomenological scenario within the MSSM is described. Sec.~\ref{analysis} is
devoted to analysing threshold corrections to the Yukawa couplings, and to
studying in what manner their unification depends on the most important
variables. In Sec.~\ref{ExamplesSec}, particular examples of points in the
MSSM parameter space with Yukawa unification are given. The impact of large
$A$-terms on flavour observables is examined in Sec.~\ref{flavSec}, whereas a
correlation with the vacuum metastability is explained in Sec.~\ref{vacSect}.

\section{The model}\label{ModelSec}

\subsection{Relevant aspects of the SUSY $\mathbf{SU(5)}$ GUT}

The SM gauge group is a subgroup of $SU(5)$. A standard embedding
of the MSSM superfields $Q$, $U$, $D$, $L$, $E$ into the 5- and
10-dimensional representations of $SU(5)$ is given by
\begin{align}
 \underbrace{(\mathbf{\bar 3},\mathbf{1},\tfrac{1}{3})}_{D} \oplus \underbrace{(\mathbf{1},\mathbf{2}, 
  -\tfrac{1}{2})}_L &= \underbrace{\mathbf{\bar 5}}_{\Psi_{\bar 5}} \\ 
\underbrace{(\mathbf{3},\mathbf{2},\tfrac{1}{6})}_Q \oplus \underbrace{(\mathbf{\bar 3},
   \mathbf{1},-\tfrac{2}{3})}_{U} \oplus \underbrace{(\mathbf{1},\mathbf{1},1)}_E &= 
\underbrace{\mathbf{10}}_{\Psi_{10}}, 
\end{align}
where the hypercharges have been displayed in the conventional SM
normalization. We are going to consider $SU(5)$ GUTs whose Yukawa terms in the
superpotential read~\cite{Dimopoulos:1981zb}
\begin{equation}
{\cal W} \ni   \Psi_{10} \mathbf{Y}^{de} \Psi_{\bar 5} H_{\bar 5} 
             + \Psi_{10} \mathbf{Y}^u \Psi_{10} H_5,
\end{equation}
where $H_{\bar 5}$ and $H_5$ are the two Higgs superfields that couple to
matter. Masses of the known fermions are thus determined  by only two
independent\linebreak $3\times 3$ matrices $\mathbf{Y}^{de}$ and
$\mathbf{Y}^{u}$. Below the gauge unification scale $M_{\rm GUT} \simeq
2\times 10^{16}\,$GeV, the model reduces to the MSSM with the superpotential
given by
\begin{equation}
{\cal W}_{MSSM}=  Q \mathbf{Y}^u U H_u +  Q  \mathbf{Y}^d D H_d +  L \mathbf{Y}^e E H_d + \mu H_d H_u.
\end{equation}
Thence, $\mathbf{Y}^d$ and $\mathbf{Y}^{e\,T}$ are equal at the
matching scale $M_{\rm GUT}$, up to a basis redefinition, and up to
threshold corrections at this scale.

Unification constraints for $\mathbf{Y}^d$ and $\mathbf{Y}^e$ take the
simplest form in the super-CKM basis where the superpotential flavour
mixing has been entirely included in $\mathbf{Y}^u$, while
$\mathbf{Y}^d$ and $\mathbf{Y}^e$ are real and diagonal. Then we just require equality
of the diagonal entries
\begin{equation}\label{yukunif}
 \mathbf{Y}^d_{ii}\stackrel{!}{=}\mathbf{Y}^e_{ii}, \hspace{20pt} i=1,2,3.
\end{equation}

Below $M_{\rm GUT}$, the relation between $\mathbf{Y}^d$ and
$\mathbf{Y}^e$ is affected by the RGE and, most importantly, by the threshold
corrections at $\mu_{\rm sp}$ that strongly depend on the soft supersymmetry
breaking terms. These terms include the gaugino masses $M_{\tilde{B}}$,
$M_{\tilde{W}}$, $M_{\tilde{g}}$, soft scalar masses 
$m^2_{\tilde{q}}$, $m^2_{\tilde{u}}$, $m^2_{\tilde{d}}$,
$m^2_{\tilde{l}}$, $m^2_{\tilde{e}}$, $m^2_{h_d}$, 
$m^2_{h_u}$, as well as the bi- and trilinear interactions of
the higgses and sfermions (squarks and sleptons)
\begin{equation}
\mathcal{L}_{soft} \ni   \tilde{q} \mathbf{A}^u \tilde{u} h_u +  \tilde{q} \mathbf{A}^d \tilde{d} h_d 
+  \tilde{l} \mathbf{A}^e \tilde{e} h_d + B\mu h_d h_u ~~+~~ {\rm h.c.}\;. \hspace{20pt}
\end{equation}
Our assumptions concerning the soft supersymmetry breaking terms are outlined in the next section.

\subsection{Choice of the parameters}

In order to approach the question of unifying Yukawa couplings by an
appropriate choice of the MSSM parameters, we assume the validity of
its RGEs up to $M_{\rm GUT}$ where the $SU(5)$ boundary
conditions are imposed.

The phenomenological motivation behind the discussed scenario within
renormalizable R-parity-conserving MSSM is to achieve Yukawa unification and
 fulfil  experimental conditions in the simplest manner, constraining as few
parameters as possible.  To independently influence the ratios
$\mathbf{Y}^d_{ii}/\mathbf{Y}^e_{ii}$ for all the three families, one needs to
adjust at least three real parameters.

Diagonal entries of the trilinear $\mathbf{A}^{de}$-terms
in the super-CKM basis (which we use throughout the article) 
can well  serve this  purpose, as they have a 
strong influence on the relevant threshold corrections. Moreover, to
obtain a correct mass of the lightest Higgs boson for given sparticle
masses, one has to adjust $\mathbf{A}^u_{33}$ that governs the 
stop mixing~\cite{Brummer:2012ns}.

Both the Higgs soft mass terms and $\tan\beta=\tfrac{v_u}{v_d}$, which we
employ to parameterize the Higgs sector, are unconstrained by the $SU(5)$
unification conditions, and can serve other phenomenological purposes.  As far
as the gaugino and the soft sfermion masses at the GUT scale are concerned, we
restrict ourselves here to the simplest choice of a common gaugino mass
$M_{1/2}$ and a universal soft mass $m_0$ for all the sfermions (but not the
Higgs doublets). Such a choice reduces the number of free parameters and makes
the analysis transparent.  However, it is by no means necessary for achieving
the Yukawa matrix unification.

In total, our scenario has 9 free parameters: $\tan\beta$, $M_{1/2}$,  $m_0$, 
$m_{h_u}$, $m_{h_d}$, $\mathbf{A}^{de}_{11}$, $\mathbf{A}^{de}_{22}$, $\mathbf{A}^{de}_{33}$, $\mathbf{A}^{u}_{33}$.

\subsection{Tools}

A standard numerical procedure that for a given parameter set leads to a full
spectrum of the MSSM can be summarized as follows. The renormalization group
equations of MSSM are solved by an iterative algorithm that interpolates
between various scales at which the parameter values are assumed. The boundary
with the SM (i.e. the scale $\mu_{\rm sp}$) is currently set by most of the
public programs to be at $M_Z$.  Such a choice has considerable
disadvantages, one of which is excluding too many parameter points
from the analysis. For instance, some fields become formally tachyonic
only well below their actual mass scale but above $M_Z$, which is still
acceptable, though most programs usually reject such points. 

Minimization of the MSSM scalar potential is performed at the scale
$M_{\rm SUSY} =\sqrt{m_{\tilde{t}_1} m_{\tilde{t}_2}}$, where the scale
dependence of the electroweak breaking conditions is relatively mild.  

A recent article~\cite{Allanach:2013cda} has shown that the contemporary
spectrum generators find only one of the potentially many models
corresponding to a given set of parameters that are specified at
multiple energy scales. In particular, it affects the cMSSM dark matter
analyses~\cite{Allanach:2013yua}. However, this fact hardly matters for
our present investigation because we only search for sample regions
in the parameter space where the Yukawa unification constraint is
satisfied.

For the purpose of the current analysis, we have modified\linebreak
{\tt SOFTSUSY~3.3.8}~\cite{Allanach:2001kg} that distinguishes itself among other
spectrum generators by possessing a technical documentation. We implemented
threshold corrections to the first and second family Yukawa
couplings as well as to the CKM matrix, given in Ref.~\cite{Crivellin:2011jt}, which {\tt SOFTSUSY} was
lacking at the moment of writing. 

\begin{table}[htbp]
\centering
\renewcommand{\arraystretch}{1.5}
\begin{tabular}{| l | l | l | l | l | l | l | l |}
\hline
$\alpha_s^{\overline{MS}} (M_Z)$ & 
$\alpha_{em}^{-1} (M_Z)$ & 
\multicolumn{2}{|c|}{$G_F\;\left[{\rm GeV}^{-2}\right]$} & 
$M_Z^{\rm pole}$  & 
$m_{e}^{\rm pole}$ &
$m_{\mu}^{\rm pole}$ \\
0.1184 &  
127.944 & 
\multicolumn{2}{|c|}{$1.16638 \times 10^{-5}$} & 
91.19 GeV &   
511 keV &
106 MeV \\
\hline
$m_{\tau}^{\rm pole}$ &
$m_u^{\overline{MS}}$ & 
$m_d^{\overline{MS}}$ &
$m_s^{\overline{MS}}$ &  
$m_c^{\overline{MS}}(m_c)$  &   
$m_b^{\overline{MS}}(m_b)$  &   
$m_t^{\rm pole}$ \\
1.777$\,$GeV &
2.3 MeV &  
4.8 MeV &
95 MeV  & 
 1.275 GeV & 
4.18$\,$GeV & 
173.5$\,$GeV \\
\hline

\end{tabular}
\caption{Standard Model parameters~\cite{Beringer:1900zz} used in
our numerical calculations.  The light ($u$, $d$, $s$) quark masses are
$\overline{\rm MS}$-renormalized at 2$\,$GeV. \label{SMinput}}
\end{table}

Our input values of the SM parameters are collected in Tab.~\ref{SMinput}.
Flavour observables are calculated with the help of {\tt
SUSY\_FLAVOR~v2.10}~\cite{Crivellin:2012jv}. This code evaluates the
renormalized MSSM Yukawa matrices and obtains the proper CKM matrix
also according to the prescriptions of Ref.~\cite{Crivellin:2011jt}, taking
the previously determined soft parameters as input. 

In our scan that delivers regions consistent with the
Yukawa unification, we use the BayesFITSv3.2~\cite{BayesFits} numerical
package that interfaces several publicly available codes. Except the
above-mentioned programs, it uses \texttt{\multinest\
v2.7}\cite{Feroz:2008xx} which allows for fast and efficient Markov Chain
Monte Carlo (MCMC) scanning according to a pre-defined likelihood function. 
For the $SU(5)$ boundary condition in Eq.~(\ref{yukunif}), we assume
a Gaussian likelihood distribution
\begin{equation}
 \like_{Yuk}=\sum_{i=1,2,3}\exp\left[-(1-Y^e_{ii}(\mgut)/Y^d_{ii}(\mgut))^2
/2\sigma_{\scriptscriptstyle\rm Yuk}^2\right],
\end{equation}
with $\sigma_{\scriptscriptstyle\rm Yuk}$ set to 10\% to allow for deviations from the exact unification condition.

\section{Analysis of threshold corrections to the Yukawa matrices 
         at $\mathbf{\mu_{\rm sp}}$}\label{analysis}

Given our choice of the GUT-scale parameters, the only source of flavour
violation at this scale is the Yukawa matrix $\mathbf{Y}^u$. Since it affects
the RGE for the remaining parameters, neither $\mathbf{Y}^d$
nor the soft terms are going to remain strictly flavour-diagonal below $M_{\rm
GUT}$. However, the corresponding flavour violation is going to be given by
the CKM matrix and remain genuinely small. While such flavour violation is
taken into account in our numerical study, we shall neglect it for simplicity
in the following discussion where large corrections to the flavour-diagonal
terms are of main interest. Within such an approximation, it is sufficient to
consider only real diagonal Yukawa matrices $\mathbf{Y}^d \equiv {\rm
diag}(Y_d, Y_s, Y_b)$ and $\mathbf{Y}^e \equiv {\rm diag}(Y_e, Y_\mu, Y_\tau)$
at all the renormalization scales.

\begin{figure}[t]
\centering
  \includegraphics[width=0.8\textwidth]{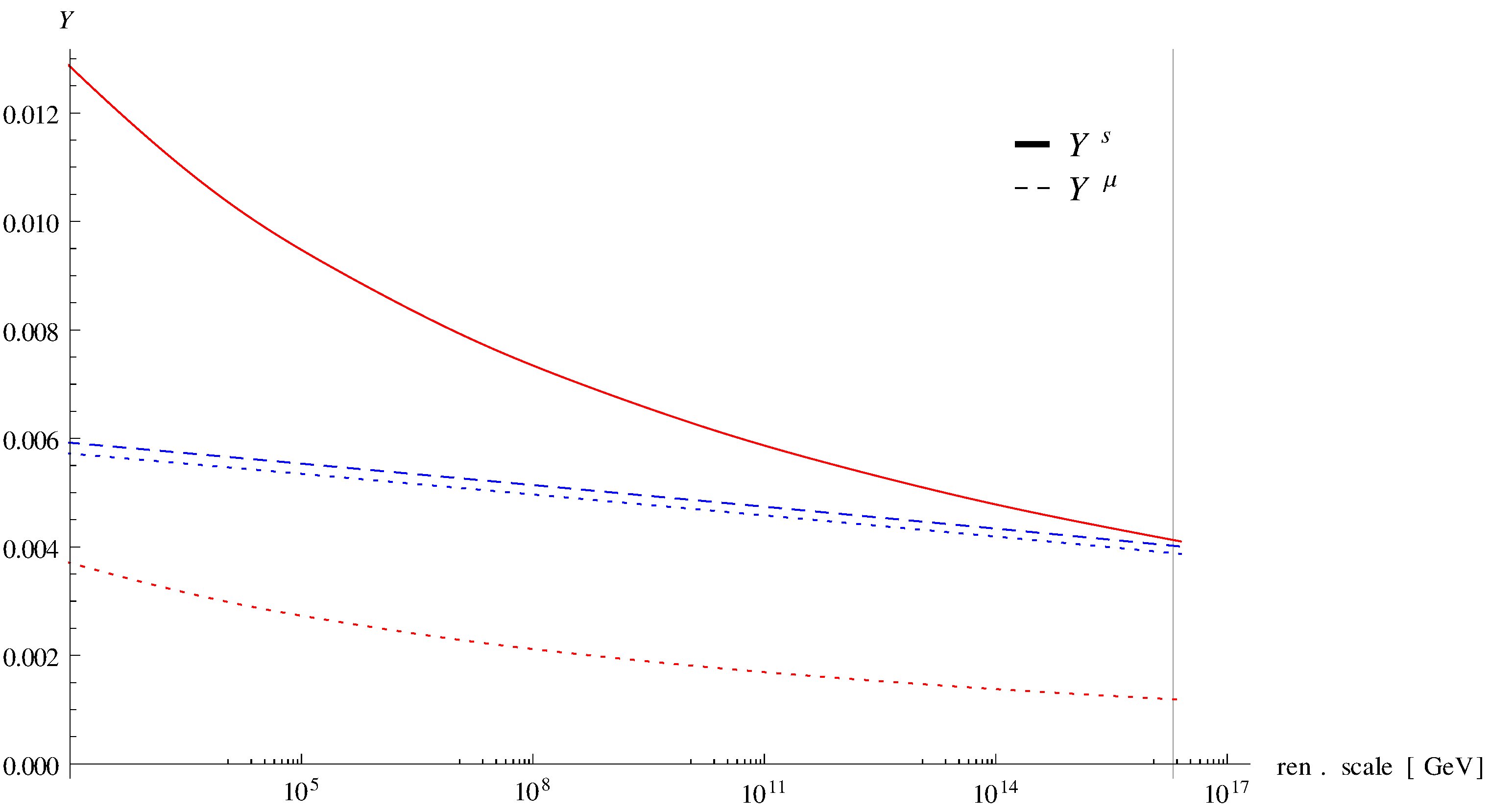}
\caption{RGE running of \Ys (red) and \Ym (blue) between 
$\mu_{\rm sp}$ and $M_{\rm GUT}$ for a sample point in the MSSM parameter
space. Dotted lines describe a situation with vanishing threshold
corrections at $\mu_{\rm sp}$. For the solid and dashed lines, the threshold
corrections have been adjusted to achieve unification at the GUT
scale.\label{YukRGErun}}
\end{figure}

As it is well known, the constraint $Y_b(M_{\rm GUT})=Y_\tau(M_{\rm GUT})$ can
be satisfied without large threshold corrections at $\mu_{\rm sp}$, at least
for moderate $\tan\beta$. On the other hand, achieving strict unification
of the other Yukawa couplings ($Y_s(M_{\rm GUT})=Y_\mu(M_{\rm
GUT})$ and $Y_d(M_{\rm GUT})=Y_e(M_{\rm GUT})$) requires the threshold
corrections to be of the same order as the leading terms. The
largest relative corrections are needed for $Y_s$, as illustrated in
Fig.~\ref{YukRGErun}.  Even in this case, the corrections are well in the
perturbative regime because the corresponding leading term is small enough
($Y_s \sim 10^{-2}$).

\SidePlotsTwot{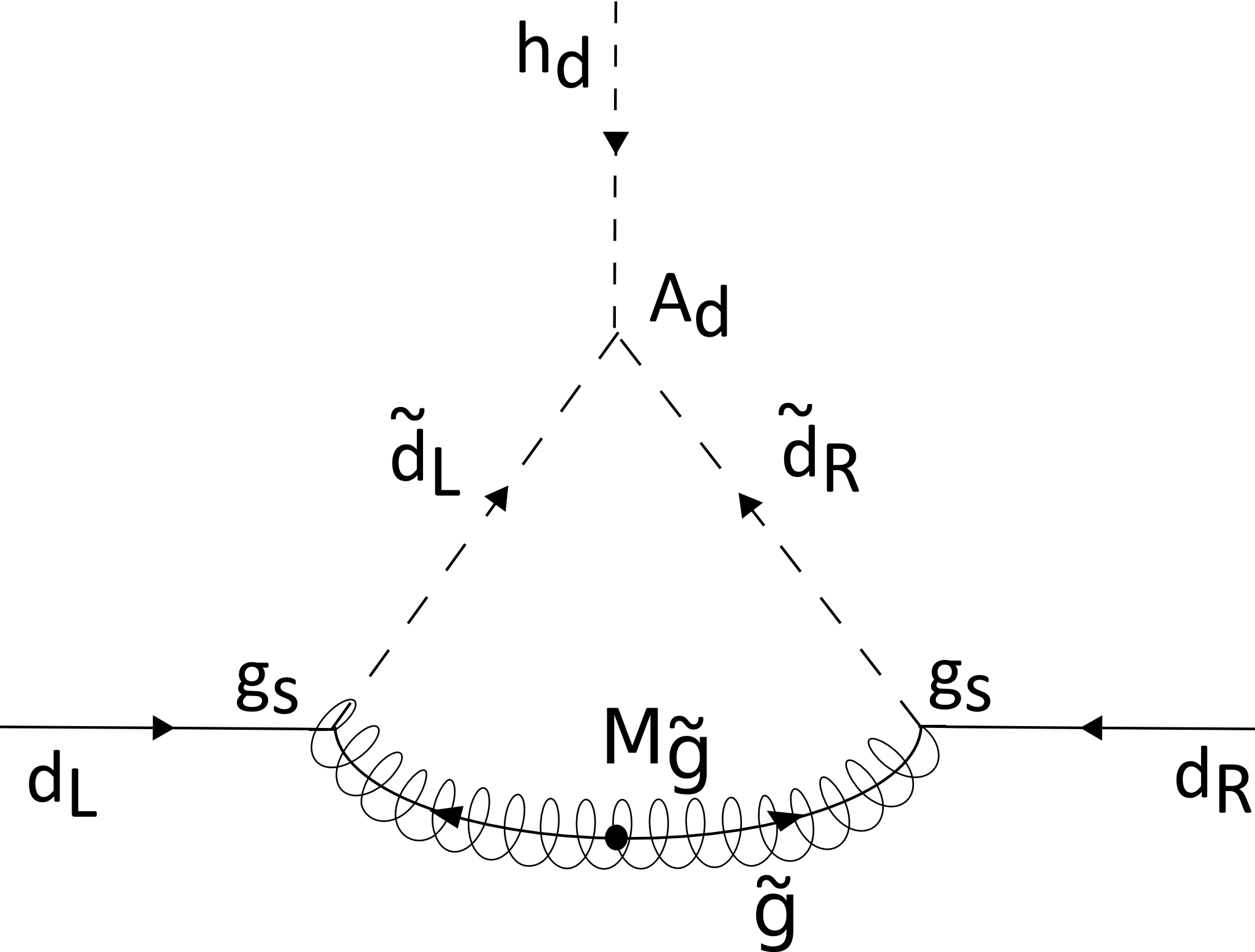}{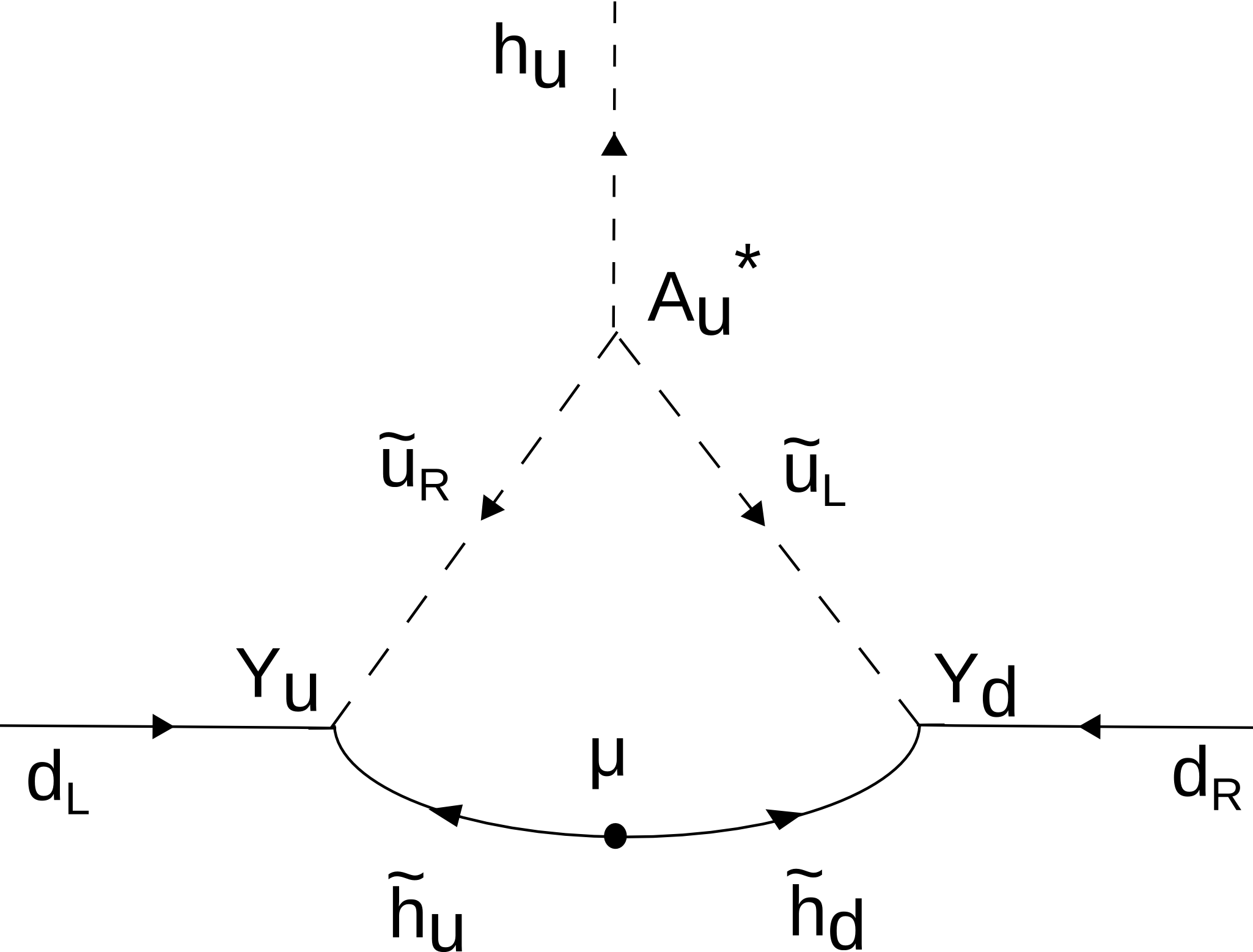}
{Examples of diagrams that describe threshold corrections to the Yukawa couplings at 
$\mu_{\rm sp}$. \label{diags}}
{0.26\textheight}      
As described in Ref.~\cite{Crivellin:2011jt}, in the SUSY-decoupling
limit, the chirality-flipping parts of the quark (lepton) self
energies $\Sigma$ are linear functions of the Yukawa couplings, with a
proportionality factor $\epsilon$ and an additive term $\Sigma_{\slashed{Y}}$
\begin{equation}
\Sigma_{ii}^{d(\ell)\,LR} \;=\;
\Sigma_{ii\,{\cancel{Y}}}^{d(\ell)\,LR} \, + \,
\epsilon_i^{d(\ell)}\,v_u\,\,\mathbf{Y}^{d(\ell)(0)}_{ii} ~+~ O(\tfrac{v^2}{M^2_{SUSY}}). 
\label{eq:epsilon_b}
\end{equation}      
Threshold corrections to the down-quark Yukawa couplings
can be enhanced by either $\tan\beta$ or large values of 
the $A$-terms. In such a case, a corrected relation between the MSSM Yukawa
couplings and the quark masses has the following approximate form:
\begin{equation}\label{TCstructEq}
\mathbf{Y}^d_{ii}=\frac{m^{d,SM}_i-\Sigma^{d,LR}_{\cancel{Y}}(\alpha_s M_{\tilde{g}} A_{ii}^{d}, m_{\tilde{q}_i},
m_{\tilde{d}_i})}{v_d [1+ \tan\beta \cdot\epsilon^d (\mu, M_{\tilde{B}}, M_{\tilde{W}}, 
M_{\tilde{g}}, m_{\tilde{q_i}}, m_{\tilde{d_i}}) ]}
\end{equation}
where $m^{d,SM}_i$ is the SM quark mass at the matching scale
$\mu_{\rm sp}$, $m_{\tilde{q}}$, $m_{\tilde{d}}$ are the
respective family squark soft masses, and $\alpha_s$ is the
strong coupling constant.

The trilinear coupling $\mathbf{A}^d_{ii}$ controls the most
significant contribution from a loop with the gluino -- see the
first diagram in Fig.~\ref{diags}. It can be used to adjust the threshold
correction and to achieve the Yukawa unification for given
values of other parameters.

In the following, we shall illustrate how the threshold corrections
to the Yukawa couplings at $\mu_{\rm sp}$
\begin{equation} \label{eq:dY}
\delta Y_{x} \equiv \frac{v_d Y_{x}^{MSSM}-m^{\rm SM}_x}{m^{\rm SM}_x},
\hspace{1cm} x = d,s,b,e,\mu,\tau,
\end{equation}
as well as the ratio $\mathbf{Y}^d_{ii}/\mathbf{Y}^e_{ii}$ at the GUT
scale depend on the most important parameters of the 
model. Using the point no.~3 in Tab.~\ref{Ex} 
(Sec.~\ref{ExamplesSec}) as a reference, we have varied only two
parameters at a time, which gives an estimate of the shape of the relation in
the vicinity of the considered point.  We concentrate on the cases of the
second and third family, as the first and second ones are qualitatively
similar.

In Figs.~\ref{m3A33}--\ref{m1A11}, we show only the points  fulfilling 
all the necessary phenomenological requirements, in particular that the
Higgs vacuum is a local minimum of the scalar potential \footnote{No scalar
tachyons appear in the spectrum.}, and that no Landau poles arise below
$M_{\rm GUT}$. White regions in the plots mean that either one of
above conditions was not fulfilled, or that {\tt SOFTSUSY} rejected the
point as its iterative algorithm had not converged.

Starting from the largest couplings, we notice that three parameters play a
crucial role in the case of bottom-tau unification: $\mathbf{A}^{de}_{33}$,
$\mu$ and $m_0$ (which for given $M_{1/2}$ governs masses of the third family
sfermions). Non-universal sfermion masses, independent for each family, could
grant additional freedom to our model. Although they are not necessary to
achieve Yukawa unification, relaxation of the universality could facilitate
finding points with even higher $\tan\beta$ than presented in the next
section.

Values of the ratio $Y_b/Y_{\tau}$ at $M_{\rm GUT}$ are presented in
Fig.~\ref{m3A33} as functions of $\mathbf{A}^{de}_{33}$ and $m_0$.  The
equality of $Y_b$ and $Y_{\tau}$ at this scale in general might demand an
adjustment of all the parameters because excluded points tightly surround the
allowed region.
\SidePlotsTwo{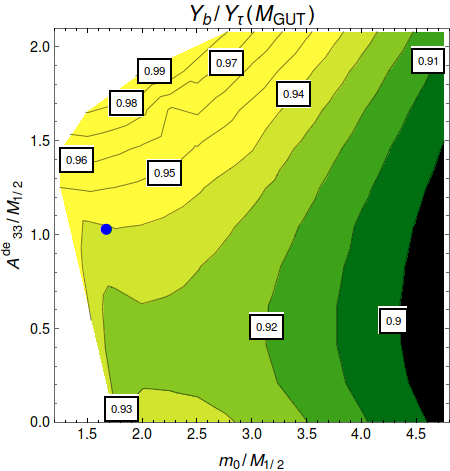}{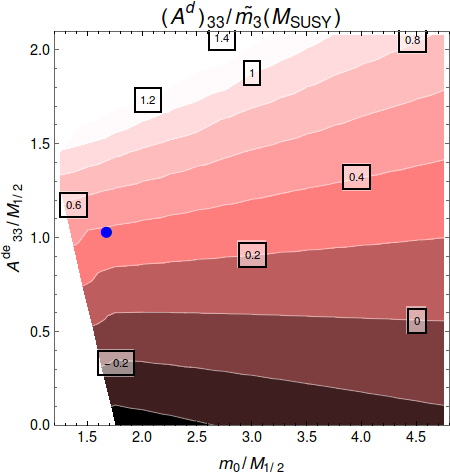} 
{Left: The ratio $Y_b/Y_{\tau}$
presented as a function of $\mathbf{A}^{de}_{33}$ and $m_0$. Right: 
The corresponding values of $\mathbf{A}^{d}_{33}/\tilde{m}_3$ at $M_{\rm
SUSY}$.  They are shown around point 3 from Tab.~\ref{Ex}
(marked by a blue dot). Both $\mathbf{A}^{de}_{33}$ and $m_0$
are normalized to $M_{1/2}$ which equals to around 815 GeV at that
point. \label{m3A33}} {0.35\textheight}
\SidePlotsTwo{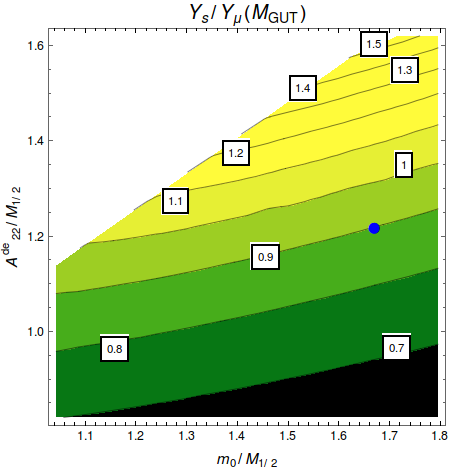}{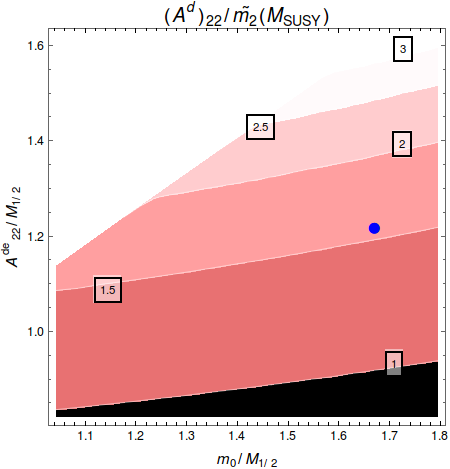}{Left: The ratio $Y_s/Y_{\mu}$
presented as a function of $\mathbf{A}^{de}_{22}$ and $m_0$. Right: 
The corresponding values of $\mathbf{A}^{d}_{22}/\tilde{m}_2$ at $M_{\rm
SUSY}$. They are shown around point 3 from Tab.~\ref{Ex}
(marked by a blue dot).
\label{m2A22}}
{0.35\textheight}
\SidePlotsTwo{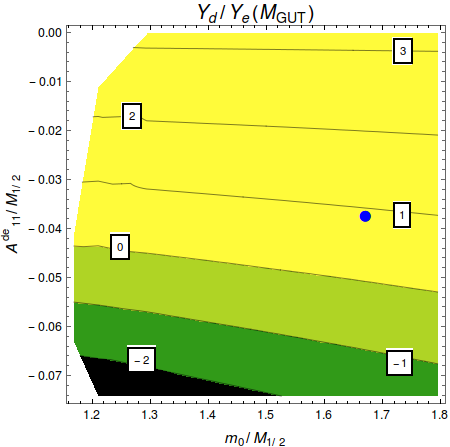}{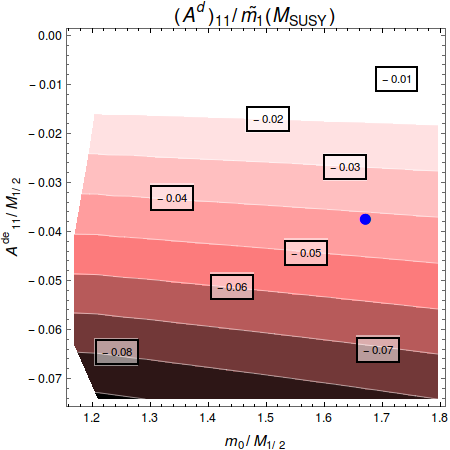}{Left: The ratio $Y_d/Y_{e}$ as
a function of $\mathbf{A}^{de}_{11}$ and $m_0$.  Right: The corresponding
values of $\mathbf{A}^{d}_{11}/\tilde{m}_1$ at $M_{\rm SUSY}$.  They are shown
around point 3 from Tab.~\ref{Ex} (marked by a blue dot).
\label{m1A11}}
{0.35\textheight}

In the second family case, unification of $Y_s$ and $Y_{\mu}$ is usually
possible by a manipulation of just one parameter, namely
$\mathbf{A}^{de}_{22}$, despite the fact that it influences both the Yukawa
couplings. For the second family, $\mu$ has little influence on the
unification in the considered region because the higgsino loop gives a much
smaller contribution, due to $m_c \ll m_t$.

The ratio $Y_s/Y_{\mu}$ at $M_{\rm GUT}$ plotted in Fig.~\ref{m2A22} against
$m_0$ and $\mathbf{A}^{de}_{22}$ shows that a large value of
$\mathbf{A}^{de}_{22}$ is required to achieve unification.  The corresponding
values of $\mathbf{A}^{d}_{22}/\tilde{m}_2$ at $M_{\rm SUSY}$ are shown in
Fig.~\ref{m2A22}. Such ratios will be relevant for our discussion of the
vacuum metastability in Sec.~\ref{vacSect}. Here, $\tilde{m}_i$ are defined by
\begin{equation}
\tilde{m}_i=\sqrt{\tfrac{m_{\tilde{q}_{i}}^2+m_{\tilde{d}_i}^2
+m_{H_d}^2}{3}}.
\label{mtildeEq}
\end{equation}

Unification of the down-quark and electron Yukawa couplings is illustrated
in Fig.~\ref{m1A11}.  It is the simplest case, because the necessary
adjustment of the respective $A$-term neither triggers any
phenomenological problems nor influences any parameters that
are relevant for other families.

\section{Regions with successful $\mathbf{SU(5)}$ Yukawa matrix unification}\label{ExamplesSec}

In Figs.~\ref{scatPlotTb}--\ref{scatPlotTb2} as well as in Tab.~\ref{Ex},
we present sample parameter-space regions and benchmark points where a proper
Yukawa matrix unification has been achieved in our setup. In selecting these
regions and points, we aimed at fulfilling the unification constraints and
reproducing the lightest Higgs particle mass (up to the theoretical
uncertainty of 3 GeV) for a broad range of $\tan\beta$. We have chosen the
sparticle masses so that the gluino is heavy enough to have evaded the current
bounds, but could possibly be detected in the second LHC phase.

Plots in Figs.~\ref{scatPlotTb}--\ref{scatPlotTb2} show points
investigated in our MCMC scans performed for three $\tan\beta$
intervals: [5,20], [15,30], and [30,45]. Different colours are used
to indicate successful Yukawa matrix unification either for all the three
families or for some of them only. We observe that Yukawa unification for all
the three generations can be achieved for a wide range of
$\tan\beta$. Generically, larger values of the $A$-terms are necessary for
larger $\tan\beta$ because the down-quark Yukawa couplings (and thus the
required threshold corrections) scale proportionally to $\tan\beta$. For this
reason, finding acceptable points for larger values of $\tan\beta$ in each
random scan required collecting much more statistics.
\SidePlotsTwoW{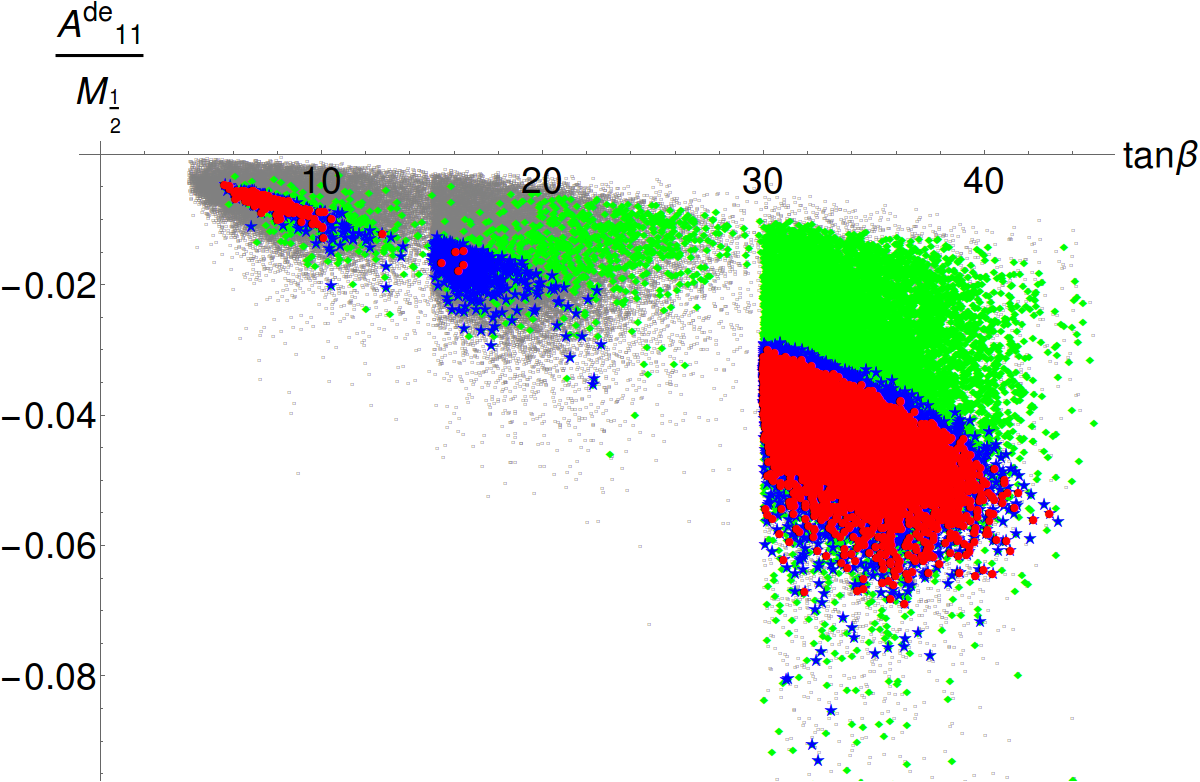}{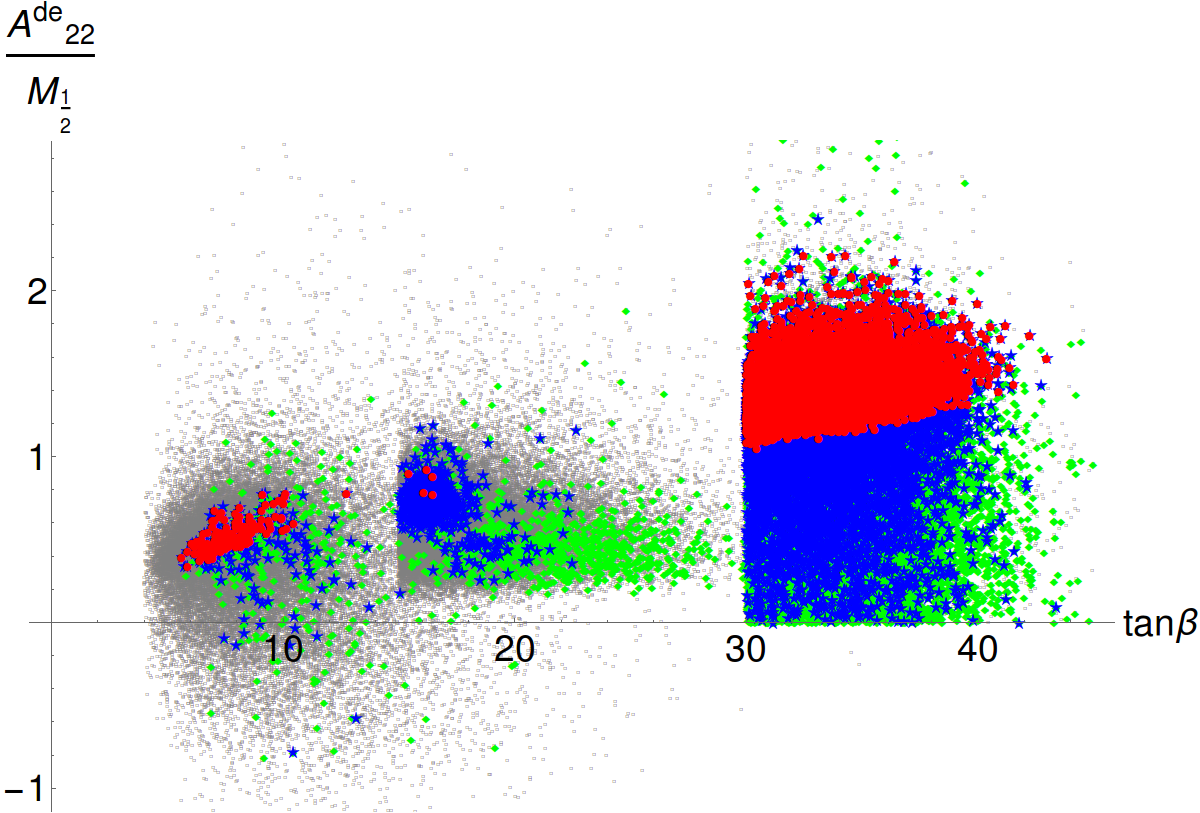}{
Left: points gathered in three of our MCMC scans (grey), shown
in the $\tan\beta \times (\mathbf{A}^{de}_{11}/M_{1/2})$ plane.  For
some of them, the respective Yukawa couplings get unified within a 10\% bound and the Higgs boson mass prediction lies in the interval $[122.5, 128.5]$ GeV:
green diamonds mark the $b$--$\tau$ unification, blue stars fulfil also the
$d$--$e$ one, while red circles include also the $s$--$\mu$
one (i.e. the full Yukawa matrices get unified). Right: the same data
projected onto the $\tan\beta \times (\mathbf{A}^{de}_{22}/M_{1/2})$ plane.
\label{scatPlotTb}}
\SidePlotsTwoW{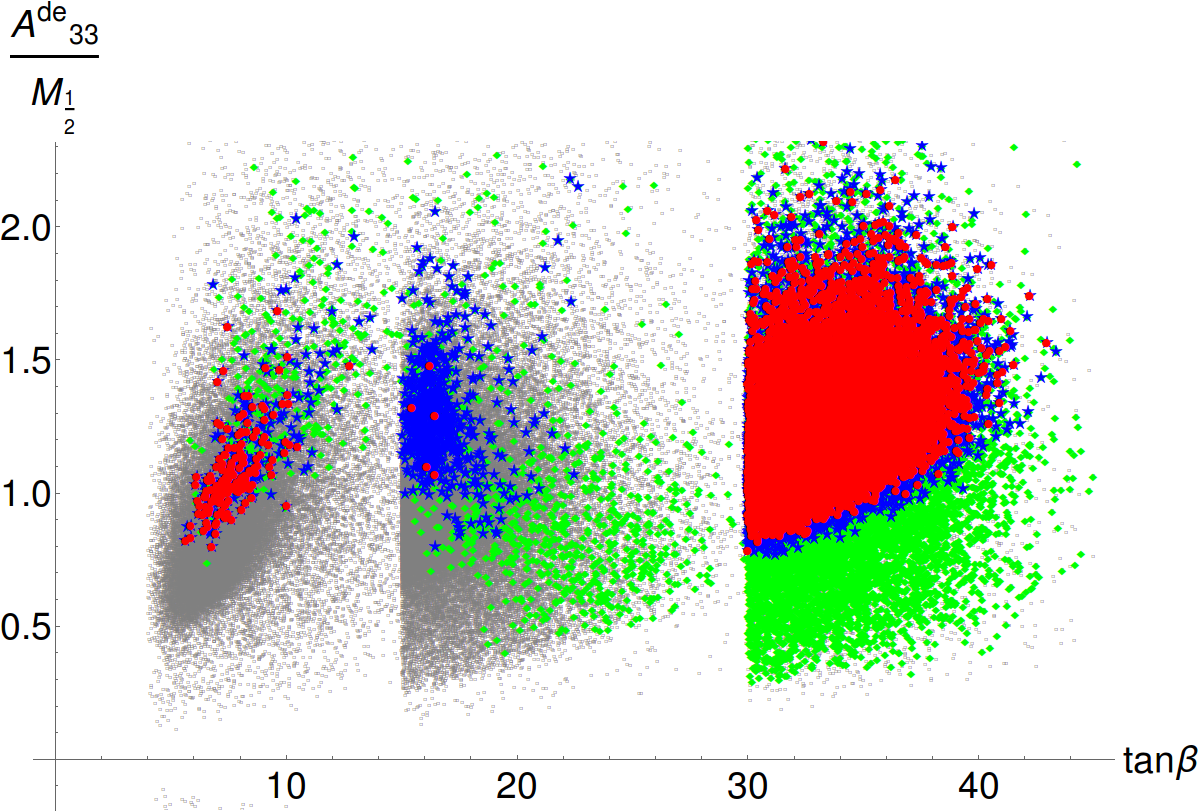}{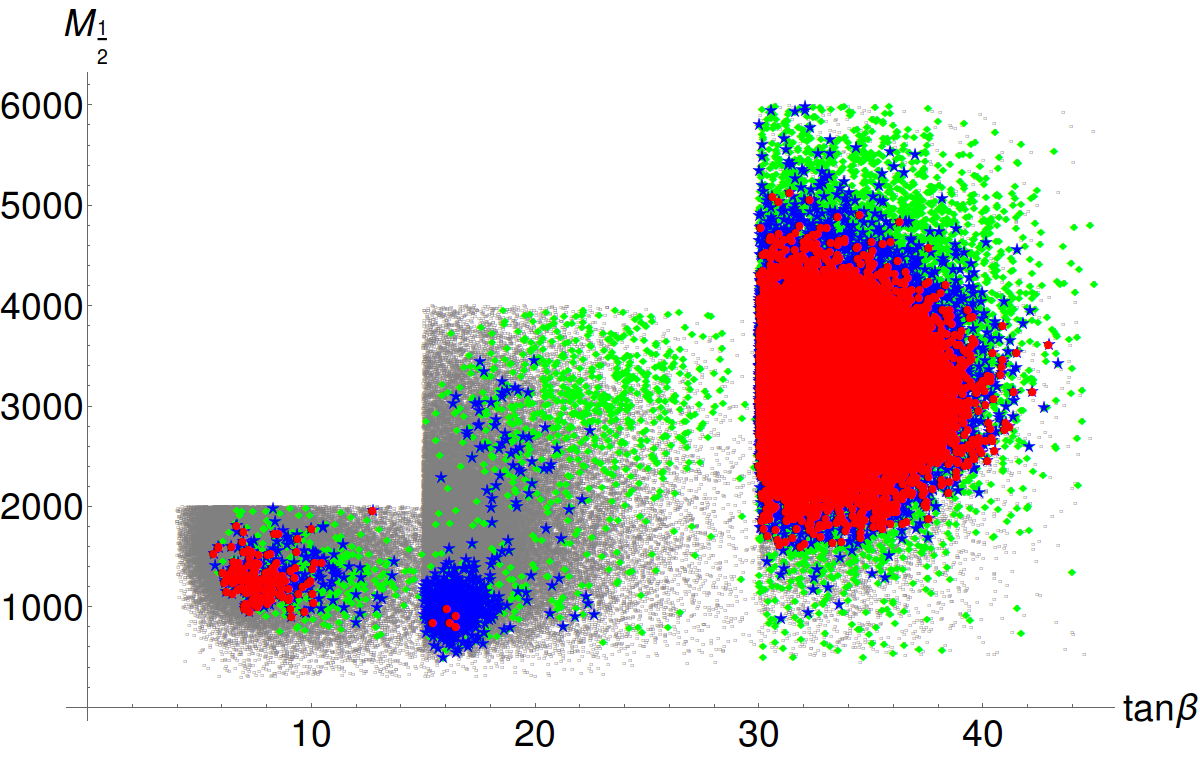}{
The same data as in Fig.~\ref{scatPlotTb} projected onto the $\tan\beta \times
(\mathbf{A}^{de}_{33}/M_{1/2})$ (left) and $\tan\beta \times M_{1/2}$ 
(right) planes.
\label{scatPlotTb2}}

Tabs.~\ref{Ex}-\ref{Outinos} contain information on the input parameters and
particle spectra in four sample points with a proper Yukawa matrix unification.
Tab.~\ref{OutEx} shows the corresponding SUSY-scale threshold 
corrections, as defined in Eq.~(\ref{eq:dY}).  In addition, we give the
GUT-scale ratios $\tfrac{Y^d_i}{Y^e_i}$ which parametrize the unification
quality. Their (small) deviations from unity determine sizes of the necessary
GUT-scale threshold corrections.  Finally, we also present the ratios
$\tfrac{m_{h_d}}{m_0}$ and $\tfrac{m_{h_u}}{m_0}$ at the GUT scale that
quantify departures from the scalar mass universality for each of the points.
\begin{table}[htbp]
\centering
\renewcommand{\arraystretch}{1.5}
\begin{tabular}{| c | c | c | c | c | c | c | c | c | c |c | c | c | c | }
\hline
 & $\tan\beta$ & $\tfrac{M_{1/2}}{{\rm GeV}}$ &  $\tfrac{m_0}{M_{1/2}}$ & 
$\tfrac{m_{h_u}}{m_0}$  & $\tfrac{m_{h_d}}{m_0}$  & $\tfrac{\mathbf{A}^{de}_{11}}{M_{1/2}}$ &
$\tfrac{\mathbf{A}^{de}_{22}}{M_{1/2}}$ & $\tfrac{\mathbf{A}^{de}_{33}}{M_{1/2}}$ 
& $\tfrac{\mathbf{A}^{u}_{33}}{M_{1/2}}$ 
 \\
\hline
1 & 7.47 & 1032.27 &  2.28 &  0.49 & 2.35 & -0.008 &  0.63 &  1.63 & -2.13 \\
2 & 16.4 & 815.081 &  1.72 &  0.37 & 2.46 & -0.017 &  0.89 &  1.30 & -1.23 \\
3 & 30.3 & 2733.63 &  1.67 &  0.19 & 1.79 & -0.037 &  1.22 &  1.04 &  1.96 \\
4 & 40.4 & 2663.45 &  1.72 &  0.26 & 2.58 & -0.055 &  1.53 &  1.65 &  1.56 \\

\hline
\end{tabular}
\caption{ Examples of points with a successful Yukawa unification. 
They are given by their defining sets of MSSM parameters: $\tan\beta$,
common gaugino mass $M_{1/2}$, common sfermion mass $m_0$, 
soft masses of Higgs doublets $m_{h_u}$ and $m_{h_d}$,
soft trilinear couplings $\mathbf{A}_{ii} (\mgut)$. \label{Ex}}
\end{table}
\begin{table}[htbp]
\centering
\renewcommand{\arraystretch}{1.5}
\begin{tabular}{| c | c | c | c | c | c | c |  }
\hline
& $\delta Y_d$ & $\delta Y_s$ & $\delta Y_b$ & $\tfrac{Y_d}{Y_e}$ & $\tfrac{Y_s}{Y_{\mu}}$ & $\tfrac{Y_b}{Y_{\tau}}$  \\
\hline
1 & -0.6 & 2.4 & 0.19 &  1.10 &  0.94 &  0.97    \\
2 & -0.66 & 2.3 & 0.089 &  0.95 &  0.91 &  0.90    \\
3 & -0.69 & 2.1 & 0.085 &  0.92 &  0.90 &  0.94    \\
4 & -0.66 & 2.1 & 0.13 &  1.01 &  0.92 &  1.08    \\

\hline
\end{tabular}
\caption{Values of the threshold corrections and other characteristics
of the points from Tab.~\ref{Ex} (see the text). \label{OutEx}}
\end{table}
\begin{table}[htbp]
\centering
\renewcommand{\arraystretch}{1.5}
\begin{tabular}{| c | c | c | c | c | c | c | c | c | c | c |}
\hline
& $m_{\tilde{s}_L}$ & $m_{\tilde{s}_R}$ & $m_{\tilde{\mu}_L}$ & $m_{\tilde{\mu}_R}$
& $m_{\tilde{t}_1}$ & $m_{\tilde{t}_2}$ & $m_{\tilde{b}_1}$ & $m_{\tilde{b}_2}$ & $m_{\tilde{\tau}_1}$ & $m_{\tilde{\tau}_2}$  \\
\hline
1 &     2031. &     1856. &     1433. &     1433. &     1328. &     1649.  &     1603.  &     1652.  &     1130.  &      1185.  \\
2 &     3049. &     2958. &     2579. &     2579. &     1542. &     2085. &     2068.  &      2154.  &     1916. &       2011.  \\
3 &  5999. &    4944.  &      3752.  &      3752.  &     5596 &    4933 &    5620  &       5698 &  3937 &      4180.  \\
4 &    5700.  &     4290.  &      3714.  &     3714.  &       4902  &      4951 &      4904.  &   3984.  &      2658. &       2658.  \\

\hline
\end{tabular}
\caption{ Masses of selected sfermions (in GeV) corresponding to the points from
Tab.~\ref{Ex}. In the case of the second generation, where the left-right mixing is negligible, mass eigenstates are labeled according to their largest interaction eigenstate component.
\label{OutSf}} \end{table}
\begin{table}[htbp]
\centering
\renewcommand{\arraystretch}{1.5}
\begin{tabular}{| c | c | c | c | c | c | c | c | c | c |}
\hline
& $m_{\tilde{g}}$ & $m_{\chi^0_1}$ & $m_{\chi^0_2}$ & $m_{\chi^0_3}$
& $m_{\chi^0_4}$ & $m_{\chi^{\pm}_1}$ & $m_{\chi^{\pm}_2}$  & $m_{A_0}$ & $\mu$ \\
\hline
1 &  2350. &   453. &   865. &  2688. &  2688. &   865. &  2689. &  5192. & -2698. \\
2 &  1858. &   350. &   670. &  1626. &  1626. &   670. &  1628. &  2981. & -1626. \\
3 &  5719. &  1223. &  2269. &  4529. &  4530. &  2270. &  4531. &  5523. & -4521. \\
4 &  5545. &  1189. &  2205. &  3690. &  3691. &  2205. &  3692. &  6958. & -3686. \\

\hline
\end{tabular}
\caption{ Masses of the gluino, neutralinos, charginos, pseudoscalar $A_0$ and the value of $\mu$ parameter (in GeV) 
corresponding to the points from Tab.~\ref{Ex}. \label{Outinos} }
\end{table}

\section{Flavour observables}\label{flavSec}

In this section, we discuss the impact of large $A$-terms on flavour
observables.  The MSSM scenario we consider does not include any
sources of flavour- and CP- violation at $M_{GUT}$ other than the CKM
matrix. Therefore, flavour off-diagonal entries of the soft terms 
remain small, as they arise solely from the RGE running.\footnote{ {\tt
SOFTSUSY 3.3.8} assumes that all the MSSM parameters are real, i.e. it neglects
the CP-violating phases. A separate numerical evaluation of the soft
term imaginary parts has been performed with the help of {\tt
SPheno 3.3.3} \cite{SPheno1,SPheno2}. No observable impact on 
CP-violating observables has been found for the MSSM parameter space points
discussed in the previous section.}

In the following, we shall illustrate how the flavour observables change when the $A$-terms 
grow from 0 to $150\%$ of the value that is necessary for 
the Yukawa unification $\mathbf{Y}^d(M_{\rm GUT})=
\mathbf{Y}^{e\,T}(M_{\rm GUT})$ to take place. Among the observables
calculable with the help of {\tt SUSY\_FLAVOR~v2.10}, only three turn
out to be significantly altered: 
\begin{center}
${\mathcal B}_\gamma \equiv {\mathcal B}(\bar B \to X_s \gamma)$,~~~
    $\overline{{\mathcal B}}_{s\mu} \equiv \overline{{\mathcal B}}(B_s \to \mu^+ \mu^-)$~~ and~~
    $\overline{{\mathcal B}}_{d\mu} \equiv \overline{{\mathcal B}}(B_d \to \mu^+ \mu^-)$.
\end{center}
Moreover, the only $A$-term component they noticeably depend on is
$\mathbf{A}^{de}_{33}$. Another important parameter to which these
observables are sensitive is $\tan\beta$.

In Fig.~\ref{bsgammaPlot}, we show the dependence of 
$\delta{\mathcal B}_\gamma \equiv ({\mathcal B}_\gamma^{\rm MSSM}-{\mathcal
B}_\gamma^{\rm SM})/{\mathcal B}_\gamma^{\rm SM}$
on $\mathbf{A}^{de}_{33}$ and
$\tan\beta$. For each example listed in Tab.~\ref{Ex} (and also for 17 other
examples), we have plotted $\delta{\mathcal B}_\gamma$ keeping all the
parameters but $\mathbf{A}^{de}_{33}$ fixed.  As one can see, SUSY
contributions in our examples can enhance ${\mathcal B}_\gamma$ by up to
$30\%$ w.r.t. the SM prediction
${\mathcal B}_\gamma^{\rm SM} = (3.15 \pm 0.23)\times 10^{-4}$~\cite{Misiak:2006zs}.  
Moreover, up to $10\%$ relative differences are observed between points with
vanishing and maximal $\mathbf{A}^{de}_{33}$. The current experimental world
average yields
${\mathcal B}_\gamma^{\rm exp}=  (3.43 \pm 0.22)\times 10^{-4}$~\cite{Amhis:2012bh}.
The observed significant variation of ${\mathcal B}_\gamma$ could lead to a
possible verification of the model once the uncertainties get reduced.
\SidePlotsTwo{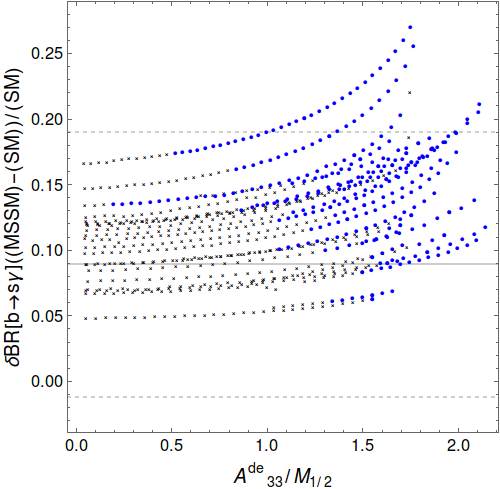}{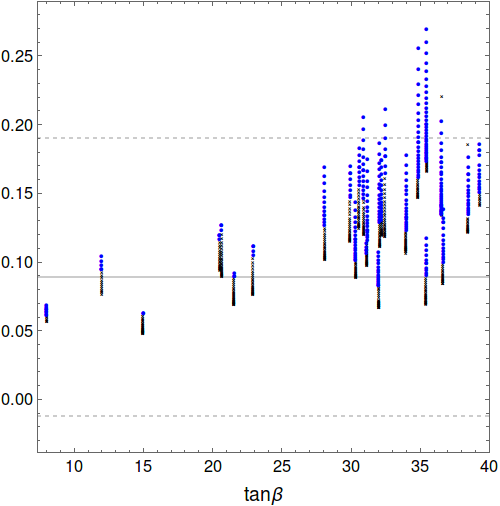}
{Dependence of $\delta{\mathcal B}_\gamma$ on
$\mathbf{A}^{de}_{33}$ and $\tan\beta$. Points fulfilling $\mathbf{Y}^d(M_{\rm
GUT})=\mathbf{Y}^{e\, T}(M_{\rm GUT})$ are marked in blue. 
The $1\sigma$ experimental error band is represented by horizontal lines. \label{bsgammaPlot}}
{0.35\textheight}
\SidePlotsTwo{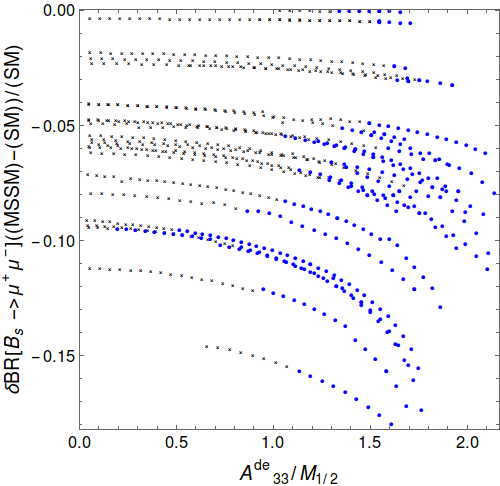}{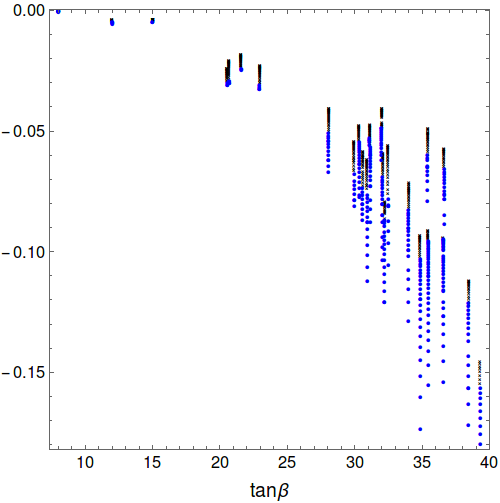}
{Dependence of $\delta\overline{\mathcal B}_{s\mu}$ on
$\mathbf{A}^{de}_{33}$ and $\tan\beta$.  For each point listed in
Tab.~\ref{Ex}, only $\mathbf{A}^{de}_{33}$ has been varied. Points fulfilling
$\mathbf{Y}^d(M_{\rm GUT})=\mathbf{Y}^{e\, T}(M_{\rm GUT})$ are marked in
blue. The results for $\delta\overline{\mathcal B}_{d\mu}$ are practically
identical.
\label{bsmumuPlot}}
{0.35\textheight}

Analogous plots for 
$\delta\overline{\mathcal B}_{s\mu} \equiv (\overline{\mathcal B}_{s\mu}^{\rm MSSM}-
\overline{\mathcal B}_{s\mu}^{\rm SM})/\overline{\mathcal B}_{s\mu}^{\rm SM}$
are shown in Fig.~\ref{bsmumuPlot}. All our sample results for
$\overline{\mathcal B}_{s\mu}^{\rm MSSM}$
fall within the $1\sigma$ band above the measurement
$\overline{\mathcal B}_{s\mu}^{\rm exp} = (2.9 \pm 0.7)\times 10^{-9}$~\cite{BsMuMuEPS,Chatrchyan:2013bka,Aaij:2013aka}
and the branching ratio can be smaller by about $15\%$ compared to the SM
prediction 
$\overline{\mathcal B}_{s\mu}^{\rm SM} = (3.65 \pm 0.23) \times 10^{-9}$~\cite{Bobeth:2013uxa}. 
As far as $\overline{{\mathcal B}}_{d\mu}$ is concerned, it undergoes an almost identical
alteration with respect to the SM. However, it remains in perfect
agreement with the present experimental result 
$\overline{\mathcal B}_{d\mu}^{\rm exp} = 
\left(3.6^{+1.6}_{-1.4}\right)\times10^{-10}$~\cite{BsMuMuEPS,Chatrchyan:2013bka,Aaij:2013aka} 
within its large uncertainties. The experimental sensitivity would need to be
improved by more than an order of magnitude to distinguish between the SM prediction
$\overline{\mathcal B}_{d\mu}^{\rm SM} = (1.06\pm 0.09)\times 10^{-10}$~\cite{Bobeth:2013uxa}
and the corresponding MSSM results for our sample points.

The three considered decays share the crucial property of
being sensitive to supersymmetric contributions even if no sources of flavour
violation beyond the CKM matrix are present. It follows from the
fact that they are all chirally suppressed in the SM.

\section{Electroweak symmetry breaking}\label{vacSect}

The MSSM contains a large number of scalar fields. In a proper analysis
of the electroweak symmetry breaking, one would need to
prove that only the neutral Higgs fields acquire non-zero values in the
global minimum of the MSSM scalar potential. However, it is well
known that there exist large regions in the MSSM parameter space where
other, deeper minima arise. At such minima, also sfermions develop
non-vanishing vacuum expectation values.

In particular, along the direction in the MSSM scalar field space where 
$$ | H_1|=|\tilde{s}_L|=|\tilde{s}_R|,$$ 
a deeper, charge and colour breaking minimum arises when $A_{s}(M_{\rm
SUSY})$ is large. Actually, all our examples in Tabs.~\ref{Ex},~\ref{OutEx}
strongly violate the stability condition~\cite{Casas:1995pd}
    $$ \frac{A_{ii}}{Y_{ii} \widetilde{m}_i} < O(1). $$
  with $\widetilde{m}_i$ defined by Eq.~(\ref{mtildeEq}). Instead, we have
    $$ \frac{A_{s}}{Y_{s} \widetilde{m}_2}(M_{\rm SUSY}) ~\sim~ 10^2. $$ 
     
However, the usual Higgs vacuum does not need to be absolutely stable.
The standard viability condition is that the its lifetime 
must be longer than the age of the Universe. According to 
Ref.~\cite{Borzumati:1999sp}, such a condition is fulfilled 
when
\begin{equation}
\frac{\mathbf{A}_{s}}{\widetilde{m}_2}<1.75\,.
\end{equation}
This requirement turns out to be satisfied in all our examples of
Yukawa unification. One can verify this by inspecting  Tab.~\ref{AEx}  where the ratios
$\tfrac{A}{\tilde{m}}$ have been presented for all the three generations.
\begin{table}[htbp]
\centering
\renewcommand{\arraystretch}{1.5}
\begin{tabular}{| c | c | c | c |  }
\hline
& $\tfrac{A_d}{\tilde{m_1}}$ &
$\tfrac{A_s}{\tilde{m_2}}$ & $\tfrac{A_b}{\tilde{m_3}}$ \\
\hline
1 & -0.0064 & 0.49 & 1.26 \\
2 & -0.016 & 0.89 & 1.01 \\
3 & -0.041 & 1.56 & 0.38 \\
4 & -0.048 & 1.54 & 0.18 \\

\hline
\end{tabular}
\caption{Values of the diagonal entries of $A$ terms at scale $M_{SUSY}$
of the points from Tab.~\ref{Ex}. \label{AEx}}
\end{table}

\section{Conclusions}

Searches for supersymmetric particles during the first LHC phase
have significantly constrained the MSSM parameter space. 
With heavier superpartners, the little hierarchy problem becomes more
difficult, but the SUSY flavour problem is rendered less severe. Thus, if
the MSSM is the proper low-energy theory, one should consider
its possibly non-trivial flavour structure. To check its consistency
with grand unification, we need to understand all the factors
involved in fulfilling the GUT boundary conditions for the
Yukawa matrices. Given only few existing analyses of the MSSM threshold
 corrections'  impact on the Yukawa matrix unification, we performed an
update that takes the recent experimental data into account.

Our article provided examples of successful $SU(5)$ Yukawa
unification that is consistent with the current experimental
bounds in a scenario where all the soft terms are
flavour-diagonal. It did come at a price. An adjustment of the down-quark
Yukawa couplings governed by the $A$-terms of the sfermion mass
size led to a conclusion that the usual Higgs vacuum becomes
metastable. However, given its long enough lifetime, such a situation is
still phenomenologically viable. 

Given the viability of the $SU(5)$ boundary conditions on MSSM Yukawa
couplings at $M_{GUT}$, it might be interesting to investigate the case of
$SO(10)$.  Unfortunately, the tiny $\tfrac{m_c}{m_t}$ ratio at the low scale
is difficult to obtain in the minimal $SO(10)$ GUT framework.  Eventually,
$SO(10)$-type unification could be achieved by employing almost complete
cancellations of threshold corrections to $Y_c$ against tree-level terms,
i.e. it would come at a price of considerable fine-tuning.

Different ways of explaining the Yukawa matrix unification are
complementary. In a general case, both the MSSM threshold corrections and the
GUT-scale higher-dimensional operators can be present. If a complete
quantitative study of a specific GUT model were to be performed, all such
options would need to be simultaneously taken into account.

We satisfied the Yukawa matrix unification constraint in possibly the simplest
manner, by adjusting just three parameters. The remaining freedom in the
choice of other soft parameters in this scenario gives it the advantage of
modularity. Such a freedom is likely to facilitate satisfying additional
phenomenological constraints (like the observed dark matter relic density) or
fitting new observables that might prove relevant for future studies.

\section{Acknowledgements}

 The author would like to thank Andreas Crivellin, Christophe Grojean,
 Gian Giudice, Kamila Kowalska, Miko{\l}aj Misiak and Ulrich Nierste for valuable discussions
 and sharing ideas.  The hospitality of CERN and the Karlsruhe Institute of
 Technology is gratefully acknowledged.  This work was supported in part by
 the Foundation for Polish Science International PhD Projects Programme
 co-financed by the EU European Regional Development Fund, by
 the Karlsruhe Institute of Technology, and by the National Science
 Centre (Poland) research project, decision DEC-2011/01/B/ST2/00438.

\end{document}